# Neutron scattering evidence for magnetic field driven abrupt magnetic and structural transitions in a phase separated manganite.

C. Yaicle[1], C. Martin[*1], Z. Jirak[2], F. Fauth[3], G. André[4],
E. Suard[5], A. Maignan[1], V. Hardy[1], R. Retoux[1],
M. Hervieu[1], S. Hébert[1], R. Raveau[1]
Ch. Simon[1], D. Saurel[4], A. Brûlet[4] and F. Bourée[4]

[1]CRISMAT, UMR6508 associée au CNRS, ISMRA, 6 Bd Ml Juin,
F-14050 CAEN Cedex, France
[2] Institute of Physics ACSR, Cukrovarnicka 10,
162 53 PRAGUE 6, Czech Republic
[3]ESRF, BP 220, 6 Rue Jules Horowitz, F-38043 Grenoble, France
[4]LLB, CEA-Saclay, F-91191 GIF-SUR-YVETTE Cedex, France
[5]ILL, 6 Rue Jules Horowitz, F-38043 Grenoble, France

Abstract :

Substitutions at the Mn-site of the charge-ordered $Pr_{0.5}Ca_{0.5}MnO_3$ manganite is an effective way to induce abrupt jumps on the magnetic field driven magnetization curve. In order to get new insights into the origin of this remarkable feature, the $Pr_{0.5}Ca_{0.5}Mn_{0.97}Ga_{0.03}O_3$ perovskite manganite has been studied by neutron diffraction, versus temperature and at 2.5K in an applied magnetic field up to 6 Tesla. A weak and complex antiferromagnetic order is found for the low temperature ground-state. Magnetic transitions, associated with structural ones, are evidenced for certain strengths of magnetic field, which gives rise to the step-like behavior corresponding to the magnetization curve. Small angle neutron scattering provides evidence for a nucleation process of micron size ferromagnetic domains which follows the magnetization behavior.

## Introduction

It is now well known that the colossal magnetoresistance (CMR) effect in some manganites is a result of strong competition between metallic ferromagnetism (FM) and insulating antiferromagnetism (AFM) [1,2]. The concept of phase separation is in fact important for compounds close to orbitally ordered state and an efficient way to induce such phenomena is to substitute a small amount of Mn by an other cation in half-doped manganites [3,4]. Depending on the parent manganite, the electronic configuration of the substituting cation and the level of substitution, various behaviors are observed [5]. Particularly interesting effects have been observed with Mn site substitutions in $Pr_{0.5}Ca_{0.5}MnO_3$, a compound which exhibits a very stable charge/orbital ordered (CO) CE-type AFM at low temperature (LT) [6,7]. Firstly, it was shown that substitutions with only few % of magnetic cations like Cr or Ru can switch this compound to a bulk FM metallic state [8]. Secondly, with non-magnetic cations like Ga or Sn at the Mn-sites (or by using a very small content of magnetic ions), amazing magnetic-field-induced step-like transitions on the magnetization curve are observed at low temperature [9]. Instead of a continuous magnetic field-induced AFM to FM transition, the FM state, at low temperature, is reached in few abrupt steps.

It has been proposed that strain in the interfacial regions between charge/orbital ordered and disordered regions could play an important role in this step-like magnetization [10]. This assumption has been proposed on the basis of T-dependent properties which are reminiscent of a martensitic-like transformation [11]. In this respect, the growth of FM regions in the AFM phase around FM nucleation centers is assisted by the application of a magnetic field. However, the different cell parameters of these phases generate strains in the interfacial regions which tend to block the AFM to FM transition. As the applied magnetic field is large enough to overcome the elastic constraints, an avalanche-like destabilization of the local stress field occurs. This abrupt growth of the FM regions is balanced by the concomitant increasing elastic energy so that a magnetization plateau can be reached. As the field increases, several magnetization jumps can thus be observed on the M *vs.* H curves collected at low enough T. Consequently with this martensitic-like scenario, a training effect for the AFM $\rightarrow$ FM transition is induced by the thermal cycling: the transition becomes more and more difficult as the T is cycled repeatedly through the transition temperature, i.e. as the AFM phase is stabilized by the thermal cycling at the expense of the FM one [12].

The aim of the present paper is to directly investigate the evolution of the magneto-structural phase separation as a function of the field, and at temperatures at which steps are observed in the magnetization curves by neutron diffraction and small-angle neutron scattering. The nature of the phase separation of the sample at low temperature can be studied by high-resolution neutron diffraction in the absence of an external magnetic field. In this respect, the $Pr_{0.5}Ca_{0.5}Mn_{0.97}Ga_{0.03}O_3$ perovskite manganite has been selected because it is a good example for the multi-step feature as has been previously reported [9, 12]. Furthermore, its magnetization can be saturated in an applied magnetic field of only 5T, and its step feature is not heavily temperature dependent for T $\leq$ 5K. Finally, from the field dependent magnetization curves it appears that each step corresponds to an irreversible sample transformation which allows the study of magnetic and structural magnetic field induced transitions by neutron diffraction after the field has been returned to zero.

In order to get information about the topology of the field-induced FM regions, the results of a small angle neutron diffraction study for $Pr_{0.5}Ca_{0.5}Mn_{0.97}Ga_{0.03}O_3$ are given. This technique is very powerful for the investigation of the shape and the size of the ferromagnetic regions in such a complex system. All the present results point towards the crucial role of the phase separation, induced by substitution at the Mn-site, in the abrupt jumps observed at low temperature in the field driven magnetization data.

**Experimental**

The synthesis conditions have been previously reported [9], i.e. starting from the stoichiometric ratio of $Pr_6O_{11}$:CaO:$MnO_2$ and $Ga_2O_3$, with several grinding and heating steps in air. In the final process, rods of 6mm diameter and several cm long were prepared, fired at 1500°C for 12 hrs and then slowly cooled (5°/min) to 800°C before being quenched to room temperature. The quality of the ceramic sample was checked by X-ray, electron diffraction and EDS (energy dispersive spectroscopy). Due to the history dependent behavior of such a compound [12], the as-prepared rods of 6mm diameter and several cm long were cut in several parts and a fresh sample was used for each measurement. As previously reported, there can be small differences in the magnetization jumps of various samples taken from the same batch, but the global shape of the M(H) curves (number of steps and approximate location of the characteristic fields) remains the same [10].

The neutron powder diffraction (NPD) study started with a characterization of this sample without an applied magnetic field, at LLB-Saclay. By using the G41 diffractometer ($\lambda$

= 2.4266Å), data were collected from room temperature (RT) to 1.5K and then RT. High resolution (HRNPD) patterns were also recorded at RT and 10K on the 3T2 diffractometer ($\lambda$ = 1.2270Å). The NPD data of the non-substituted manganite $Pr_{0.5}Ca_{0.5}MnO_3$, characterized using the same equipment, are used as a reference for the CE-type AFM charge-ordered phase [14]. After this first set of characterizations of the crystallographic and magnetic state as a function of temperature, in the absence of external magnetic field, a field-dependent study was performed at 2.5K. This temperature is chosen to aid comparison with the magnetization curves [9, 12]. These NPD patterns were collected on the D2B diffractometer at ILL-Grenoble ($\lambda$=1.5940Å) with an applied magnetic field perpendicular to the beam. Before progressive application of the magnetic field, it was first verified that the 0T data agreed well with the G41 (1.5K) and 3T2 (10K) data. The successive field values used for recording the patterns are 0, 1, 1.5, 2, 2.5, 3, 0, 3.5, 4.5, 0, 5.5 and 0T. The field is progressively increased at different values and a few measurements at 0T were intercalated to verify that the field does not induce a bias to the NPD data. The G41, 3T2 and D2B data were refined using the Fullprof program [15]. For all of the NPD experiments, the sample used was in the form of a rod, thus hindering possible grain reorientation under the magnetic field. Small angle neutron scattering (SANS) was carried out on the PAXY spectrometer at the Orphée reactor (LLB-Saclay) with a neutron wavelength of 4.5Å and a distance between the sample and the detector of 5.62m. This allows the exploration of the scattering vector range 0.01 < Q < 0.1 $Å^{-1}$. The Ga-substituted polycrystalline sample was cooled down to 2K, i.e. a temperature close to T = 2.5K as used for NPD. The magnetic field was applied as a horizontal plane perpendicular to the neutron beam. The procedure for field application is the same as previously mentionned : the field is set to a given value, and the scattering data recorded, the field is then re-set to zero and the scattering measured again. For comparison with a field-induced transition, the same kind of experiments were also carried out at 10K. Absolute signals, I(Q) in $cm^{-1}$, were extracted from the determination of the neutron flux in the incident beam [13] following the procedure used in [16].

## Results and discussions

### *Neutron diffraction versus temperature*

The RT structures of $Pr_{0.5}Ca_{0.5}Mn_{0.97}Ga_{0.03}O_3$ and $Pr_{0.5}Ca_{0.5}MnO_3$ are refined in the space group Pnma with similar cell parameters ($a_p\sqrt{2}$, $2a_p$, $a_p\sqrt{2}$) but these compounds evolve differently as a function of temperature (Fig 1). Two sets of AFM peaks are observed for $Pr_{0.5}Ca_{0.5}Mn_{0.97}Ga_{0.03}O_3$ at low temperatures. The first set corresponds to the CE-type phase, characteristic of the parent compound [7, 14]. This appears around ≈80K upon decreasing temperature and disappears around ≈100K upon increasing T, which is at much lower temperatures than in $Pr_{0.5}Ca_{0.5}MnO_3$ ($T_N$=170 K). The remaining magnetic lines are attributed to the pseudo-CE structure with $T_N$ ≈60K, both upon increasing and decreasing the temperature. This type of AFM is not observed in $Pr_{0.5}Ca_{0.5}MnO_3$ but is common in $Mn^{3+}$ richer compounds, such as $Pr_{0.6}Ca_{0.4}MnO_3$ [6,7]. It should be recalled that these different magnetic structures, CE-type and pseudo CE-type AFM, are associated with different structural distortions. At low temperature, 10K, the HRNPD pattern is thus refined with two cells : the majority phase (around 60%), retaining the RT orthorhombic distortion, is associated with pseudo-CE AFM, and the second one, characterized by an additional strong compression of the *b* axis, corresponds to the CE structure. It should be noted that all of the ratios given in this paper are those of the crystallographic phases, and do not come from magnetic parameters.These refinements lead, for the CE structure, to magnetic moments of ≈ 1.0 and ≈ 1.7 $\mu_B$ for the $Mn^{3+}$ and $Mn^{4+}$ sites, respectively. For the pseudo-CE structure, an ordered moment is only found for the $Mn^{4+}$ site, with a value of ≈ 1 $\mu_B$. The coherence lengths are refined to ≈ 1200 Å and ≈ 700 Å for the regular (associated with pseudo-CE) and distorted (corresponding to CE) phases, respectively. These results clearly evidence a poor magnetic order, with refined Mn magnetic moments far from the theoretical values in both AFM structures (less than half the expected value for the sample). This observation can be understood considering that, due to local perturbations induced by the impurity on the Mn-sites, the charge/orbital ordered phases are largely susceptible to orbital order defects like discommensurations, which are observed even for $Mn^{3+}$:$Mn^{4+}$ (1:1) systems with the CE type AFM ground state [17]. A high density of these defects disorders spins in the $Mn^{3+}$ sublattice, whilst the ordering in the $Mn^{4+}$ sublattice is much less affected. This explanation can be used also for the phase with lower lattice distortion (associated with pseudo-CE), where the same kind of magnetic moment array, observed in the FM coupled CE planes, also leads to similar defects.

The results of the RT and 10K HRNPD refinements are given in Table I, with the crystallographic parameters and selected inter-atomic distances and angles. The corresponding 10K refined pattern is also given (Fig.2). The MnO lattice distortion is clearly different in the two LT structures, as shown by the Mn-O distances. The $MnO_6$ octahedra are slightly flattened (1.933Å) with a regular basal plane (1.945 and 1.943Å) for the main phase whereas a more compressed octahedron (1.921Å) which is strongly distorted in the basal plane (1.934 and 1.969Å) is observed for the other phase. Such a difference could be responsible for significant stress in the matrix. The LT magnetic state of $Pr_{0.5}Ca_{0.5}Mn_{0.97}Ga_{0.03}O_3$ is thus close to the one observed for $Pr_{0.5}Ca_{0.5}Mn_{0.95}Al_{0.05}O_3$, with a somewhat different CE/pseudo-CE AFM ratio and higher values of ordered Mn moments for the Al-sample [8]. These results can also be compared with those obtained for $Pr_{0.5}Ca_{0.5}Mn_{0.95}Cr_{0.05}O_3$ which exhibits a different behavior with bulk FM at low temperatures, with a moment of 3$\mu_B$ refined per Mn [8]. The LT lattice parameters of $Pr_{0.5}Ca_{0.5}MnO_3$ and substituted compounds with trivalent cations on the Mn-sites are summarized in Table II. The maximal distortion is observed for the CE-AFM state of $Pr_{0.5}Ca_{0.5}MnO_3$ and the minimal distortion for the FM state of $Pr_{0.5}Ca_{0.5}Mn_{0.95}Cr_{0.05}O_3$. For the Al and Ga substituted samples, a biphasic state is obtained, in which the lattice distortions of coexisting phases are thought to be important for the observation of magnetization steps. To be discussed more precisely, the crystallographic results have to be completed with synchrotron data and electron microscopy *vs.* T observations, which is beyond the scope of this paper that is mainly devoted to the magnetic aspect of these samples. The results given here were obtained by refining the structures in the small Pnma ($a_p\sqrt{2}$, $2a_p$, $a_p\sqrt{2}$) cells, that is, without taking into account the supercells attributed to the charge ordering. The thermal dependence of the cell parameters derived from G41 data refinements, however, reveals interesting features (Fig. 3). Firstly, it is clearly shown that the lattice distortions develop progressively over a large temperature range. Secondly, as expected in the aforementioned martensitic-like scenario, one observes that the structural transition is not perfectly reversible with temperature: the unique Pnma structure refined at room temperature, after the decrease and the subsequent increase of T, exhibits cell parameters slightly different from the first set of parameters (initial sample), as shown in Fig. 3. This increase in the lattice volume with the thermal cycling has been verified by successive X-ray measurements from RT to 90K. This effect could be linked with previous reports concerning the decrease of the maximal magnetization value due to thermal cycling [9, 12]. This suggests the stabilization of the AFM phase by this type of cycling through the temperature transitions.

***Neutron diffraction versus applied magnetic field***

After the characterization of the magnetic and crystallographic state without applied magnetic field, a field-dependent study of the $Pr_{0.5}Ca_{0.5}Mn_{0.97}Ga_{0.03}O_3$ sample was performed at 2.5K on D2B following the field cycling previously described in the experimental section. Additional measurements were also carried out in zero field after 3, 4.5 and 5.5T. The results of this series of measurements are summarized in Fig. 4. The most important results are, firstly, the appearance of FM at ≈ 2.5T, at the expense of the pseudo-CE AFM that quickly disappears for ≈ 3.5T and, secondly, the increase of the FM associated with a slight decrease in intensity of the peaks characteristic of the CE-AFM for H> 3.5T. Thus, in the same way that steps are observed in the M(H) curves (Fig. 5a), the present study also reveals clear jumps on the magnetic peak intensity *vs.* H curves (Fig. 5b) for magnetic field values in rather good agreement with those of the magnetization steps in the M(H) curves. The evolution of the cell parameters *vs.* H also shows clear signatures of the M steps feature (Fig.6). The large distortion of the Pnma structure associated to the CE-AFM varies only slightly, in contrast to the less distorted Pnma structure that shows a clear evolution in the range of the pseudo-CE AFM and FM coexistence, i.e. for 2.5 ≤ H ≤3.5T, and finally becomes more regular for H ≥ 3.5T when pseudo-CE disappears. Consequently, the magnetic changes induced by the magnetic field are associated with structural changes. From the point of view of internal stress, it might be important that the difference (estimated by the [$b\sqrt{2}/(a+c)$]) between the distortion of "distorted and regular" cells increases by applying a magnetic field but the fraction of the distorted phase decreases somewhat.

Even if we have to be very cautious concerning the refinement of magnetic part of the neutron diffraction patterns measured with magnetic field, the comparison between the results obtained at 5.5T and those obtained without a magnetic field, after having being subjected to 5.5T, is interesting. As expected from the $M(H)_{2.5K}$ curve, the magnetic field induced transition is irreversible, as evidenced by the similarity of the patterns recorded in 0T after 5.5T and in 5.5T which both differ from the initial state (only a small decrease of the FM peak is observed). It must be emphasized that in both cases (5.5 and 0 after 5.5T), no more pseudo-CE peaks are observable whereas the CE peaks persist which points towards the robustness of the CE-type phase. The AFM to FM transition is also confirmed by the observation of a field-induced FM component on the diffraction patterns, associated with a more regular Pnma structure. The refinements show that this phase expands up to ≈ 80% of the matrix, with large magnetic moments of $2.6\mu_B$/Mn at 0T after 5.5T .

***Small angle neutron diffraction versus applied magnetic field***

At zero field (10K or 2.5K), the small angle scattering is isotropic (Fig. 7a) and presents, in the whole Q range, the $Q^{-4}$ dependence characteristic of the Porod regime [18], due to the scattering by sharp interfaces. Since under these conditions, the sample does not exhibit any FM, the SANS signal arises from interfaces between the compound and the vacuum. Using Porod's equation, $Q^4I(Q) = 12\pi\Delta\rho^2/R$ (where $\Delta\rho$ is the contrast length between the compound and the vacuum [18], one can extract a typical grain size of about 2R ≈5μm, assuming grains are isotropic (note that this value is in agreement with the size determined by scanning electron microscopy).

Application of a strong enough magnetic field creates ferromagnetism in the sample, as shown in the magnetization curves (Fig. 5a). In presence of aligned ferromagnetism, the SANS signal becomes anisotropic (Fig. 7b), since magnetic scattering is observed perpendicular to the moment (to the applied magnetic field, horizontal on Fig. 7b). By the difference between the SANS signal perpendicular and parallel to the field, one can extract its magnetic part. We have applied this procedure at each temperature and magnetic field. A typical Q dependence of the magnetic signal is shown on Fig. 8, showing a $Q^4$ dependence. Using Porod's equation for this magnetic part $Q^4 I_m(Q) = 4\pi\Delta\rho_m^2 S_m/V$, a typical magnetic specific surface can be extracted. In this formula, V is the sample volume and $S_m$ is the surface of the interface between the ferromagnetic and non-ferromagnetic parts. Note that $\Delta\rho_m$ is the contrast between the ferromagnetic and the non ferromagnetic part of the sample, proportional to the $3.5\mu_B$ magnetization multiplied by $0.27\ 10^{-12}$ cm/$\mu_B$ - the scattering length corresponding to $1\mu_B$- divided by the volume occupied by a formula unit corresponding to one Mn.

At 10K, as shown in Fig. 5c, one can see that the specific surface $S_m/V$ of the ferromagnetic domains increases as the ferromagnetic fraction $\phi$ increases (deduced from the magnetization curve Fig. 5a). More precisely, for small M and $\phi$ values, $S_m/V$ is roughly proportional to $\phi$, suggesting that the ferromagnetic domains are spheric and not connected. Indeed if the spheres of radius R are not connected, the ratio between the surface of the ferromagnetic spheres $S_m = 4\pi NR^2$ and the total sample volume $V = N4\pi R^3/3\phi$ is $S_m/V = 3\phi/R$ (N is the number of ferromagnetic spheres in the sample). The corresponding average size of the spheres is ≈ 4000Å. The growth of the droplet size would have provided a smaller variation of $S_m/V$ versus $\phi$. One can also notice that the surface increases in field of up to 6T and starts to decrease at 6.6T. The latter result is due to the fact that the droplets are closed packed at this ferromagnetic fraction ($\phi$ about 80%) and that their collapse decreases the interface.

At 2K, precise quantitative analysis of SANS curves is difficult, since the amplitudes of the magnetization jumps (a fortiori the ferromagnetic fraction) cannot be perfectly reproduced, even by using a fresh sample for each measurement. However, when comparing Fig. 5a (2K, M(H) curve) and c (2K, $S_m/V$(H) curve), it is clear that similar jumps are observed in both experiments. Looking at the magnetization curves, we see that the ferromagnetic fractions for a given applied magnetic field are comparable at both 2 and 10K. The specific surfaces are nonetheless 2.5 times larger at 2K. The specific surface at 2K corresponds to droplets of about 1µm. This is still 5 times smaller than the grain size (5 µm).

Finally, the comparison of SANS data at 2K and 10K shows that the phase transition occurs in both cases at micrometric scales, giving sharp interfaces between ferromagnetic and non ferromagnetic parts of the sample, even if the process is continuous at 10K and step-like at 2K.

## *Conclusion*

This neutron diffraction study evidences a step-like behavior *vs.* H for $Pr_{0.5}Ca_{0.5}Mn_{0.97}Ga_{0.03}O_3$. "Structural and magnetic" jumps are observed, at 2.5K, in accordance with the M(H) measurements. It is clear that the metamagnetic transition takes place in several stages with intermediate magnetic arrangements. The intensity of the different magnetic peaks jumps for particular values of the applied magnetic field. Such a behavior is closely connected to the LT poor magnetic order that is induced by Ga substitution.

In particular, at low temperature, 3% of Ga substitution at the Mn-site induces an antiferromagnetic state, not well established, made up of two kinds of AFM, CE-type and pseudo CE-type. Additionally, no long-range FM can be found from neutron diffraction, in agreement with the low FM fraction (~ 1%) previously estimated from M(H) measurements. It should be recalled that low refined magnetic moments are obtained for Mn which may indicate that a part of the sample is only short-range ordered.

By applying a magnetic field at 2.5K, the steps observed on the M(H) curves are explained by the discontinuous disappearance of the pseudo-CE type phase and the abrupt increase of the FM phase content. Each step in the $M(H)_{2.5K}$ curve can thus be associated with an abrupt growth of the FM fraction at the expense of the AFM phases, the pseudo CE AFM disappearing more easily. Let us note that the CE structure still persists up to 5.5T. These transitions have also clear signatures on the SANS experiments, which indicate that the field-induced step-like growths of FM fraction essentially take place via nucleation of micrometric spherical domains.

## Figure captions

Fig. 1: NPD ($\lambda$ = 2.4266Å) patterns of $Pr_{0.5}Ca_{0.5}Mn_{0.97}Ga_{0.03}O_3$ recorded at RT and 1.5K compared with that of $Pr_{0.5}Ca_{0.5}MnO_3$ 1.5K. The indexed star indicates the main super-structure peak due to charge ordering. The solid arrows correspond to the main CE type AFM and the others to the pseudo-CE AFM.

Fig. 2: 10K-3T2 ($\lambda$ = 1.2270Å) NPD (calculated and experimental) patterns. The Bragg ticks are, from top to bottom, for "regular" Pnma, pseudo-CE AFM, "distorted" Pnma and CE AFM.

Fig. 3: Cell parameters, refined in the Pnma space group, versus temperature (from NPD data, $\lambda$ = 2.4266Å). The full and hollow symbols correspond to the decreasing and increasing temperature respectively. The size of the symbols correspond to the maximal error.

Fig. 4: D2B patterns ($\lambda$ = 1.5940Å) recorded at 2.5K under different applied magnetic fields (indicated in the diagram).

Fig. 5: (a) Magnetization curve M(H) of $Pr_{0.5}Ca_{0.5}Mn_{0.97}Ga_{0.03}O_3$ collected (in zero field cooling) at 2.5 and 10K.

(b) Intensity of the magnetic peaks *vs.* H corresponding to the NPD patterns of Fig. 4.

(c) Specific surface of the interfaces from magnetic small angle neutron scattering at 2K and 10K as function of magnetic field.

In (b) and (c) the horizontal lines are only a guide to the eye, since only few points are available in comparison to (a).

Fig. 6: Magnetic field dependence of the lattice parameters of both Pnma phases observed for $Pr_{0.5}Ca_{0.5}Mn_{0.97}Ga_{0.03}O_3$ at 2.5K (from D2B data, $\lambda$ = 1.5940Å). The phase labelled "2" (corresponding to the solid lines that are only a guide to the eye) remains associated with CE-AFM whereas the "1" (and dotted lines) corresponds to pseudo CE- AFM for $H \leq 2T$ and to FM for $H \geq 3.5T$.

Fig. 7: Small angle scattering plane at 2K, without (a) and with (b) magnetic field, showing the ferromagnetic nature of the signal induced by 6T.

Fig. 8: Intensity of the magnetic small angle scattering at 6T and 2K showing the $Q^{-4}$ dependence.

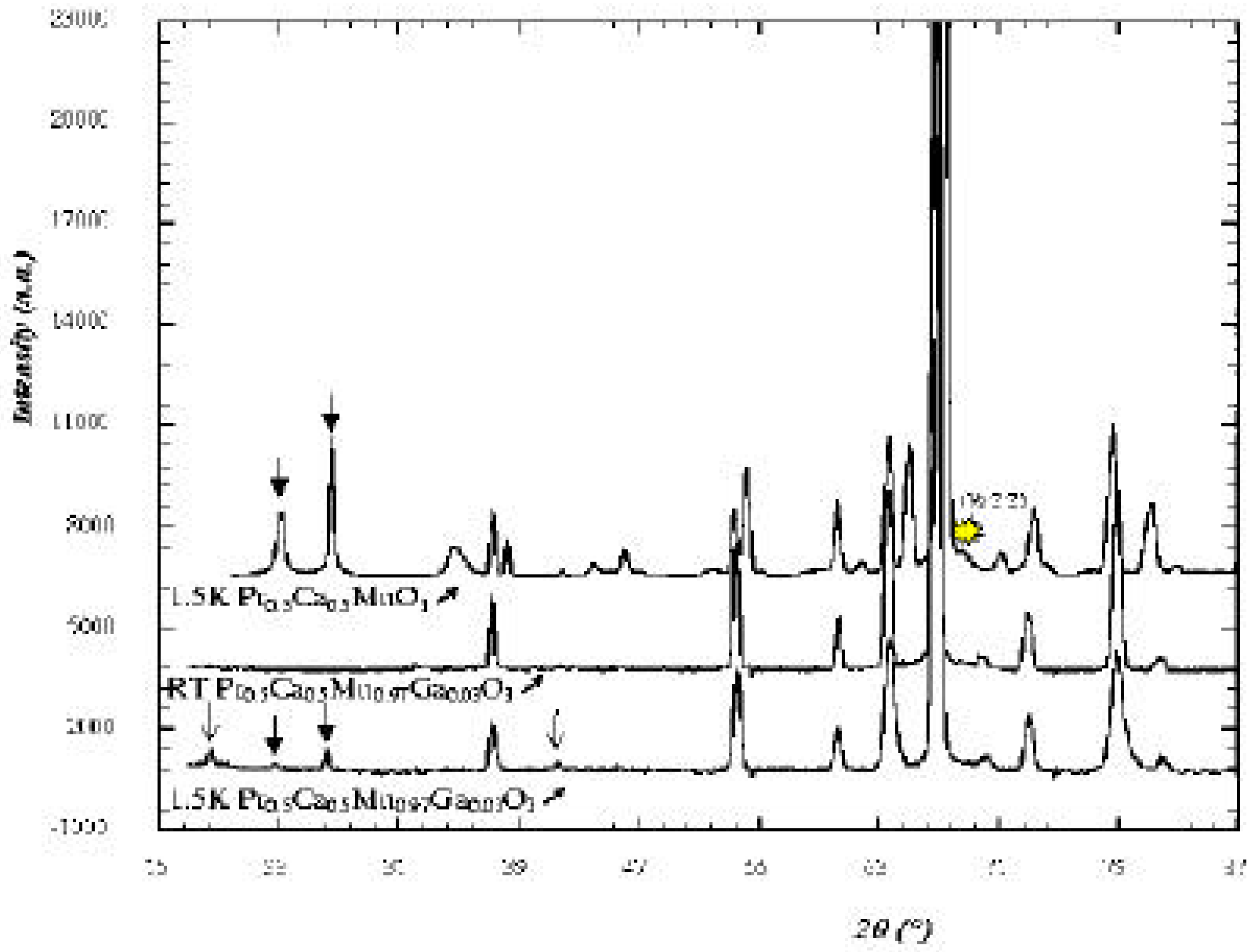

**figure1**

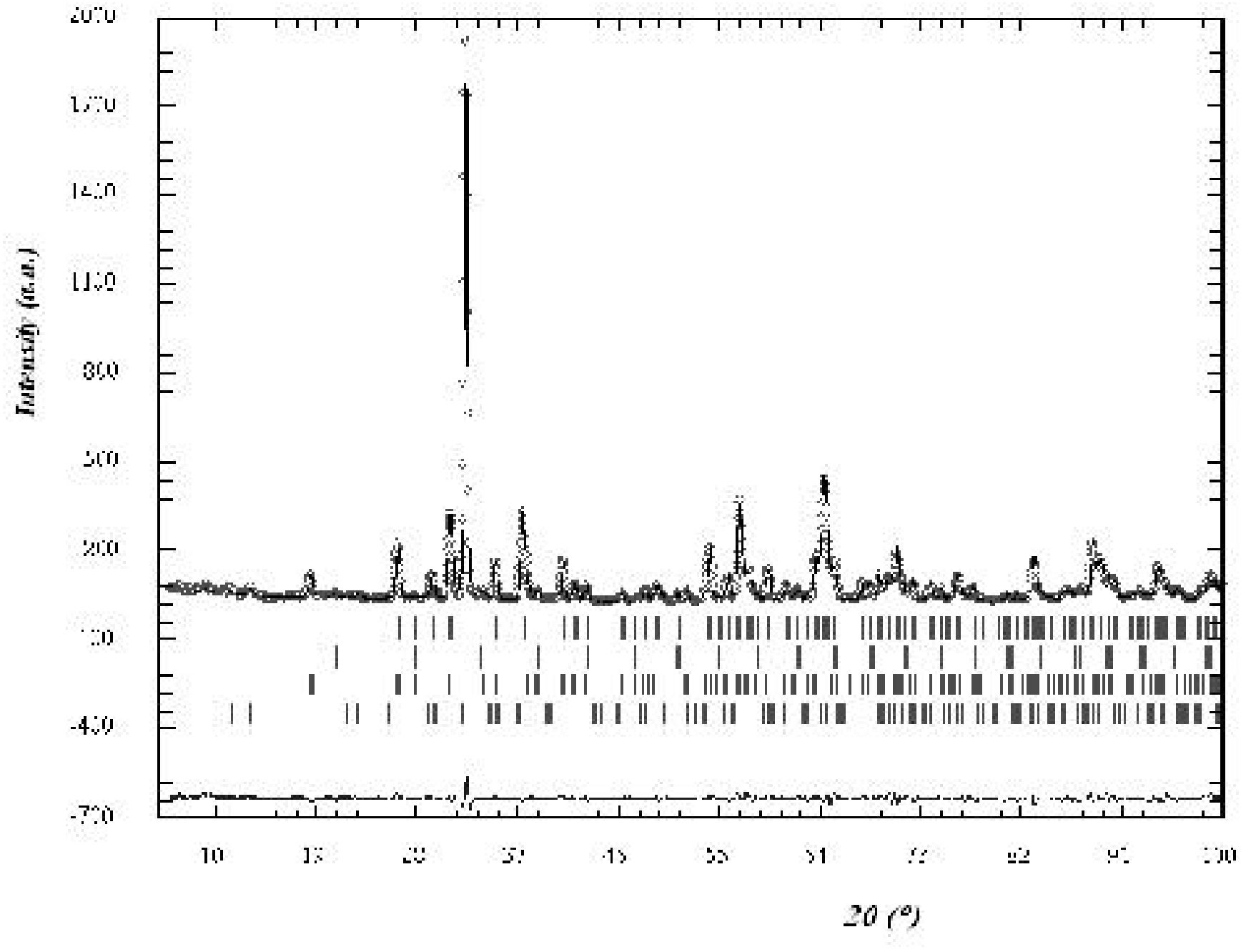

**figure2**

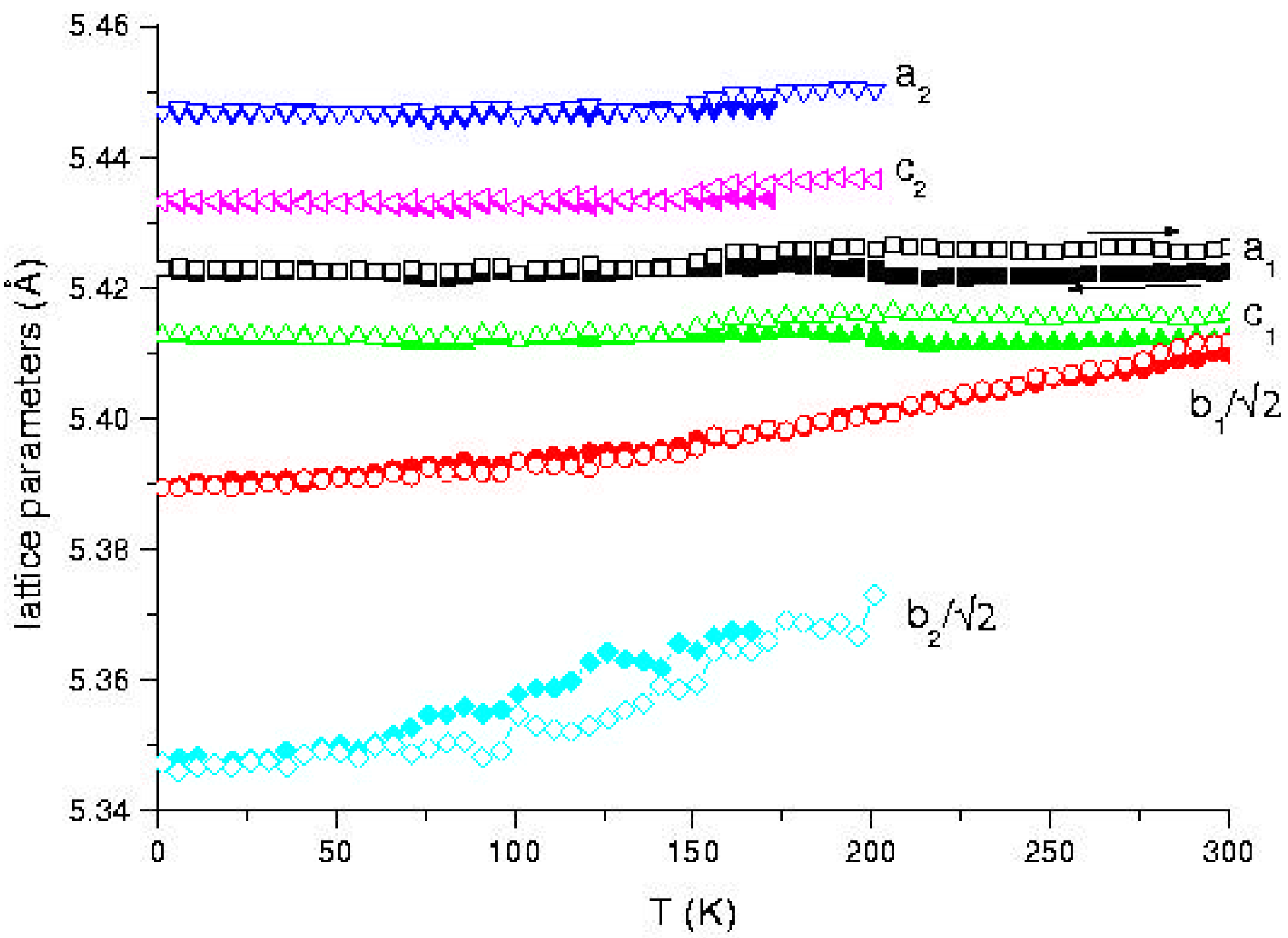

**figure 3**

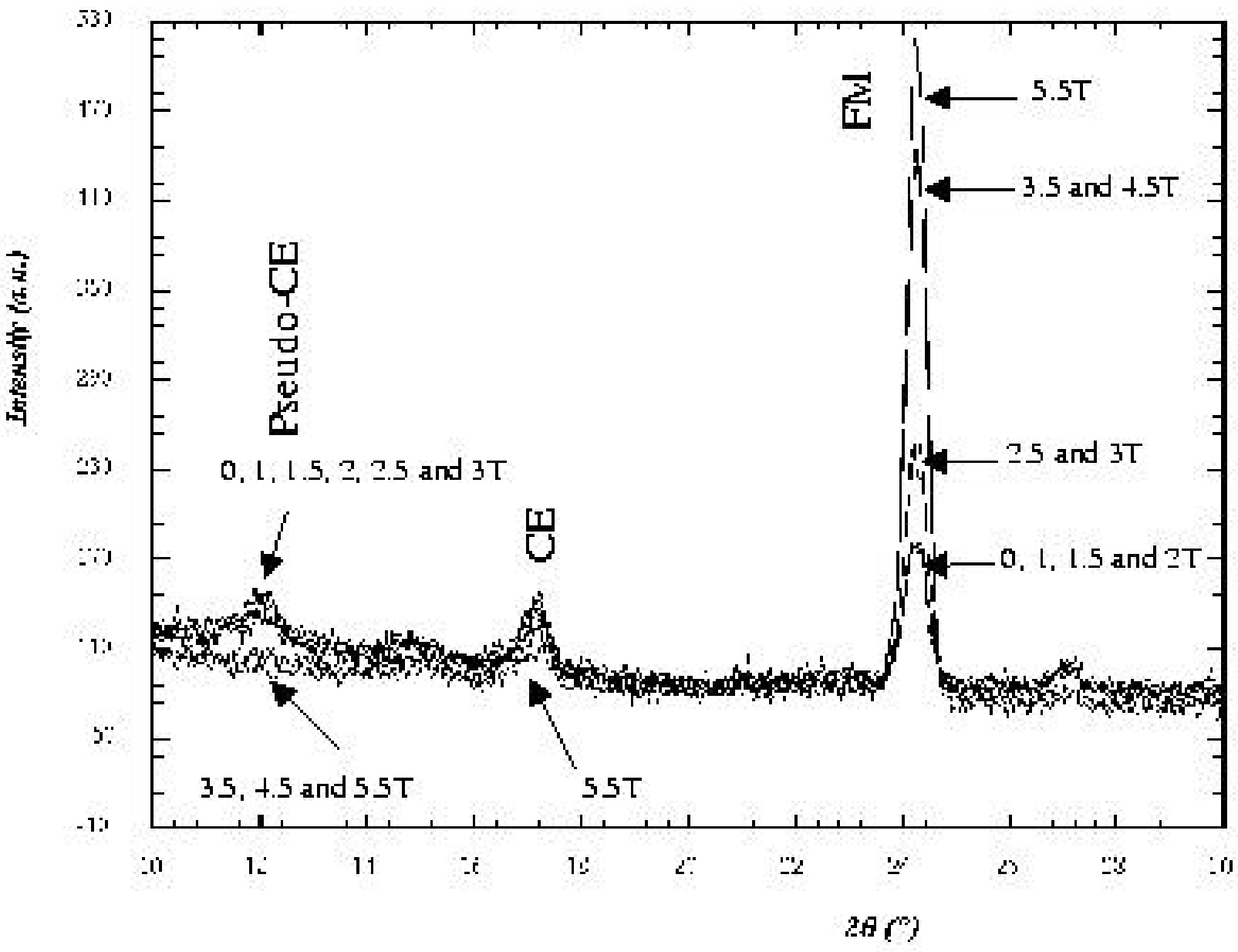

**Figure 4**

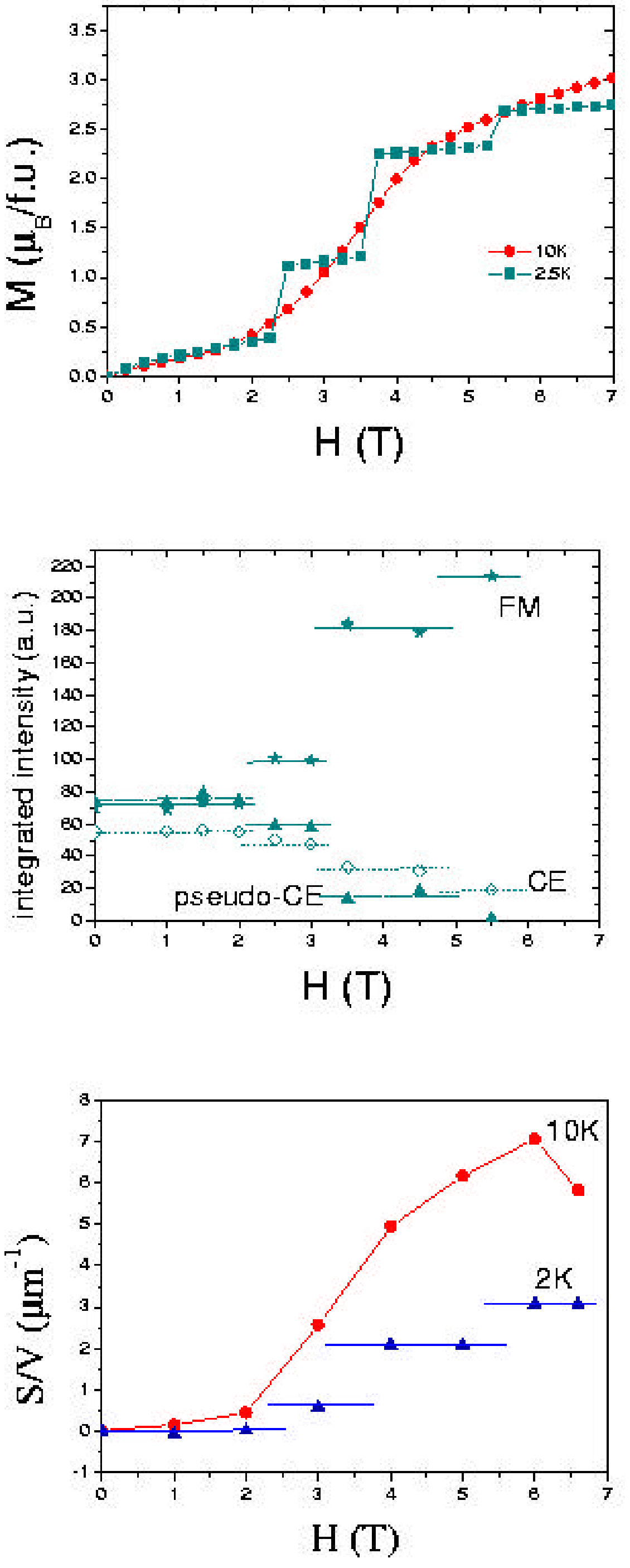

**Figure 5**

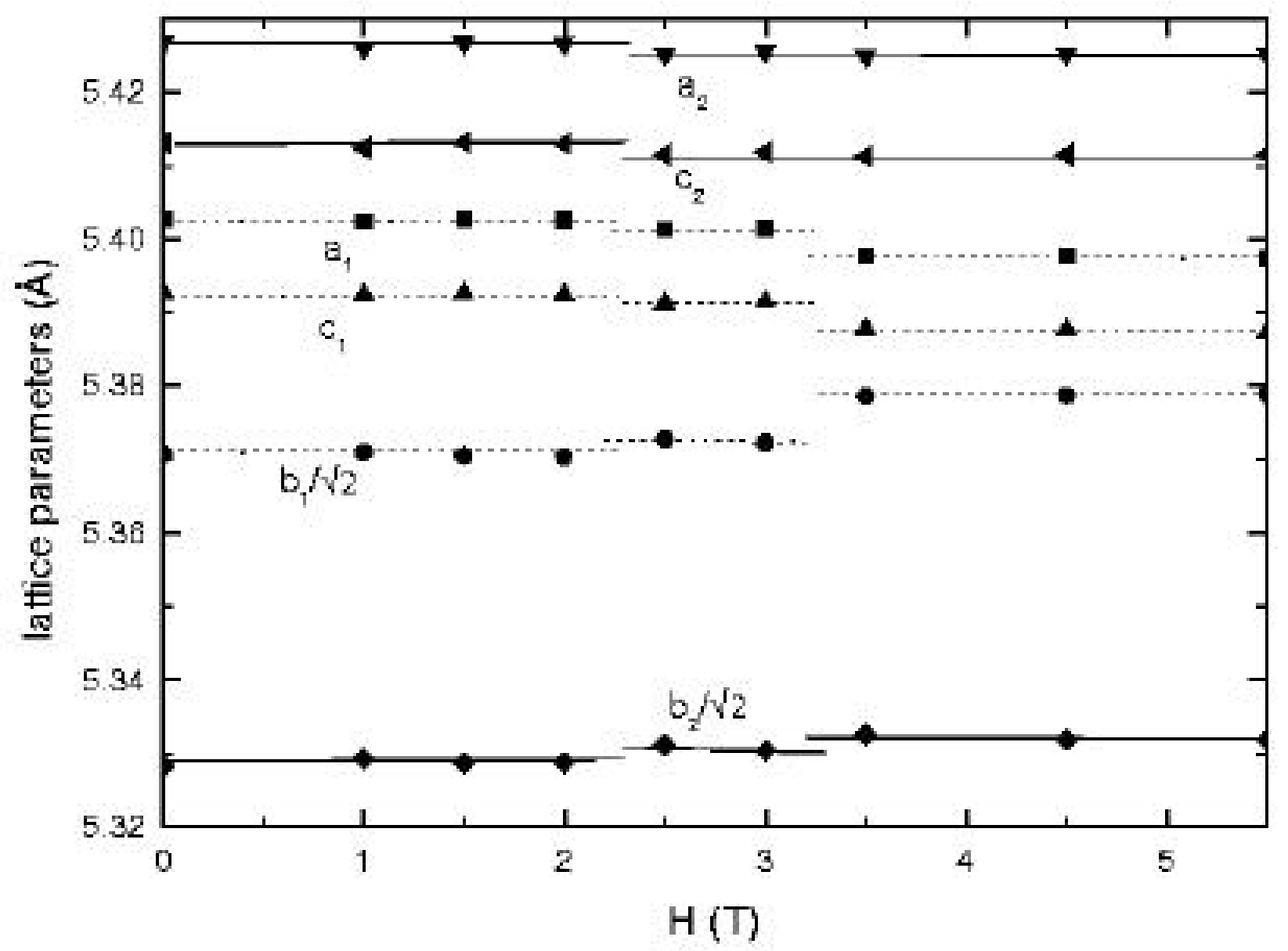

**Figure 6**

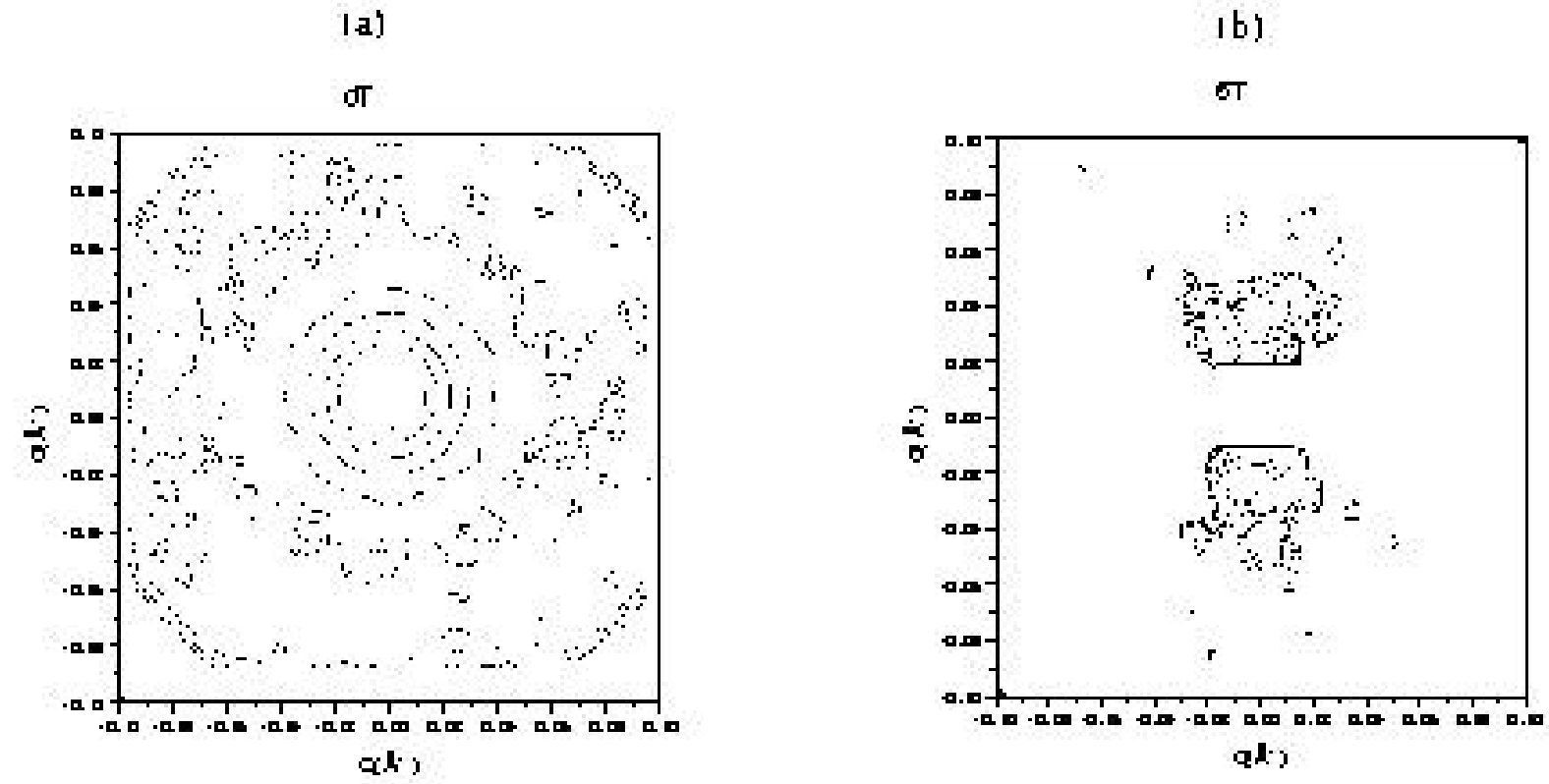

**Figure 7**

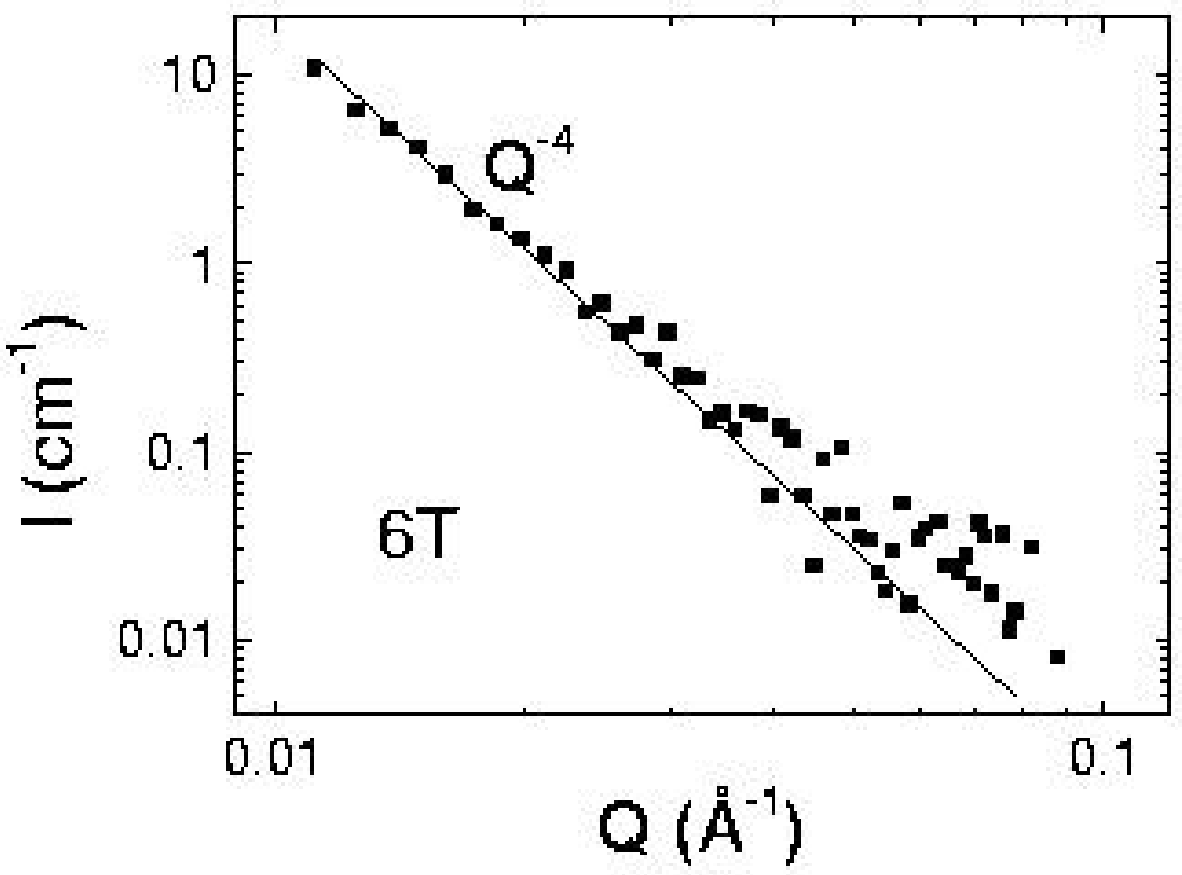

**Figure 8**

| Temperature | 300K | 10K | |
|---|---|---|---|
| | | phase 1 (pseudo CE) 60% | Phase 2 (CE) 40% |
| Space group | *Pnma* | *Pnma* | *Pnma* |
| a (Å) | 5.40115(7) | 5.4006(1) | 5.4246(1) |
| b (Å) | 7.62031(9) | 7.5932(2) | 7.5386(3) |
| c (Å) | 5.39365(6) | 5.3905(1) | 5.4109(1) |
| volume ($Å^3$) | 221.995(5) | 221.06(1) | 221.27(1) |
| Pr, Ca site 4c | | | |
| x | 0.0300(3) | 0.0337(6) | 0.031(1) |
| y | 0.25 | 0.25 | 0.25 |
| z | -0.0073(5) | -0.007(1) | -0.010(1) |
| B ($Å^2$) | 0.61(2) | 0.69(6) | 0.4(1) |
| Mn, Ga site 4c | | | |
| x | 0 | 0 | 0 |
| y | 0 | 0 | 0 |
| z | 0.5 | 0.5 | 0.5 |
| B ($Å^2$) | 0.23(2) | 0.6(1) | 0.6 |
| O(1) site 4c | | | |
| x | 0.4893(3) | 0.4889(7) | 0.513(1) |
| y | 0.25 | 0.25 | 0.25 |
| z | 0.0666(4) | 0.0663(7) | 0.067(1) |
| B ($Å^2$) | 0.79(4) | 0.82(5) | 0.44(9) |
| O(2) site 8d | | | |
| x | 0.2851(3) | 0.2838(5) | 0.2853(8) |
| y | 0.0345(1) | 0.0358(2) | 0.0363(4) |
| z | -0.2847(2) | -0.2843(5) | -0.280(1) |
| B ($Å^2$) | 0.72(2) | 1.01(4) | 0.84(8) |
| Rwp (%) | 2.96 | 3.68 | 5.01 |
| $\chi^2$ | 2.11 | 4.88 | |
| Mn-O(1) (Å) *2 | 1.940(04) | 1.933(07) | 1.921(1) |
| Mn-O(2) (Å) *2 | 1.947(1) | 1.945(3) | 1.969(5) |
| Mn-O(2) (Å) *2 | 1.943(1) | 1.943(3) | 1.934(5) |
| Mn-O(1)-Mn (°) | 158.37(7) | 158.4(1) | 157.6(2) |
| Mn-O(2)-Mn (°) | 157.80(7) | 157.7(1) | 157.9(2) |

**Table I :** Crystallographic parameters, selected interatomic distances and angles of $Pr_{0.5}Ca_{0.5}Mn_{0.97}Ga_{0.03}O_3$ at 300 and 10K (from 3T2 NPD data).

| | a (Å) | b (Å) | b/√2 (Å) | c (Å) | *d* |
|---|---|---|---|---|---|
| $Pr_{0.5}Ca_{0.5}MnO_3$ | 5.4348 | 7.4819 | 5.2905 | 5.4335 | 0.9736 |
| $Pr_{0.5}Ca_{0.5}Mn_{0.95}Al_{0.05}O_3$ | 5.4227<br>5.3940 | 7.5032<br>7.5784 | 5.3056<br>5.3587 | 5.4219<br>5.3940 | 0.9785<br>0.9935 |
| $Pr_{0.5}Ca_{0.5}Mn_{0.97}Ga_{0.03}O_3$ | 5.4246<br>5.4006 | 7.5386<br>7.5932 | 5.3306<br>5.3692 | 5.4109<br>5.3905 | 0.9839<br>0.9951 |
| $Pr_{0.5}Ca_{0.5}Mn_{0.95}Cr_{0.05}O_3$ | 5.4016 | 7.6082 | 5.3798 | 5.3875 | 0.9973 |

**Table II:** Comparison of the 10K cell parameters (Pnma space group) and lattice distortion [d = b√2/(a+c)] of different Mn-site substituted $Pr_{0.5}Ca_{0.5}Mn_{1-y}M_yO_3$ compounds (from 3T2 data, λ = 1.2251 Å). [(Ga, y = 0.03) this work, (Cr, y = 0.05) [19], (Al, y = 0.05) [8] and parent sample (y = 0) [14]]

| | a (Å) | b (Å) | b/√2 (Å) | c (Å) | *d* |
|---|---|---|---|---|---|
| $Pr_{0.5}Ca_{0.5}MnO_3$ | 5.4348 | 7.4819 | 5.2905 | 5.4335 | 0.9736 |
| $Pr_{0.5}Ca_{0.5}Mn_{0.95}Al_{0.05}O_3$ | 5.4227<br>5.3940 | 7.5032<br>7.5784 | 5.3056<br>5.3587 | 5.4219<br>5.3940 | 0.9785<br>0.9935 |
| $Pr_{0.5}Ca_{0.5}Mn_{0.97}Ga_{0.03}O_3$ | 5.4246<br>5.4006 | 7.5386<br>7.5932 | 5.3306<br>5.3692 | 5.4109<br>5.3905 | 0.9839<br>0.9951 |
| $Pr_{0.5}Ca_{0.5}Mn_{0.95}Cr_{0.05}O_3$ | 5.4016 | 7.6082 | 5.3798 | 5.3875 | 0.9973 |

**Table II:** Comparison of the 10K cell parameters (Pnma space group) and lattice distortion [d = b√2/(a+c)] of different Mn-site substituted $Pr_{0.5}Ca_{0.5}Mn_{1-y}M_yO_3$ compounds (from 3T2 data, λ = 1.2251 Å). [(Ga, y = 0.03) this work, (Cr, y = 0.05) [19], (Al, y = 0.05) [8] and parent sample (y = 0) [14]]